# Comparison of three numerical stabilization techniques of viscoelastic flows: vortex shedding behind a confined cylinder

Sai Peng [1,2] and Peng Yu [1,2,3*]

[1]Guangdong Provincial Key Laboratory of Turbulence Research and Applications, Department of Mechanics and Aerospace Engineering, Southern University of Science and Technology, Shenzhen, China

[2]Shenzhen Key Laboratory of Complex Aerospace Flows, Department of Mechanics and Aerospace Engineering, Southern University of Science and Technology, Shenzhen, 518055, China

[3]Center for Complex Flows and Soft Matter Research, Southern University of Science and Technology, Shenzhen, 518055, China

**Abstract**

In this study, the OpenFOAM platform, based on the finite volume method, is applied to investigate the two-dimensional viscoelastic flow past a circular cylinder. The FENE-P model, which considers the bounded elongation of polymer molecules, is chosen to describe the elastic constitutive relationship of the polymer solution. The maximum molecular chain lengths of $L$ = 10, 50, 100, and 200 are considered, which describe the molecular conformation characteristics of the polymer solution. To improve the numerical instability of the viscoelastic flow simulation, three different methods, i.e., the traditional method (*Td*) with the addition of artificial viscosity, the logarithmic reconstruction method (Log), and the square root tensor method (Sqrt), are evaluated. The results show that the artificial viscosity has a little effect on the accuracy for the simulation with a small molecular chain length ($L$ = 10). However, for long molecular chain lengths such as $L$ = 100 and $L$ = 200, the addition of artificial dissipation tends to overestimate the drag, which indicates that special caution is needed to incorporate the artificial dissipation in the simulation. Moreover, the logarithmic reconstruction method shows a strong grid-dependent characteristics, which may produce unphysical results.

## 1. Introduction

Addition of soluble polymer into water can greatly change rheological properties of fluid, thus affecting flow behavior, such as suppressing flow instability or turbulence and reducing wall friction drag (White *et al.*, 2008). Therefore, it is regarded as a potential flow control method. For example, Xiong *et al.* (2019) proposed a strategy of adding solvable polymer into water to inhibit vortex induced vibration. The method uses the principle of polymer additives to suppress the flow instability. Although there are many application backgrounds, it is still unclear how the polymer additives are related to turbulence (White *et al.*, 2008). The change of flow after polymer addition is mainly due to the introduction of extra-elastic stress. However, it is extremely difficult to model and calculate the elastic stress, especially in the case of high Weissenbeg number(*Wi*), which is commonly called the well-known high Weissenberg number problem (HWNP, Alves *et*

***Corresponding Author:** yup6@sustech.edu.cn

*al.*, 2021).

The research on viscoelastic channel or pipeline flow, is more abundant than flow over a blunt body. However, compared with the channel or pipeline flow, viscoelastic flow over a blunt body has its specificity. When flowing around a blunt body, there is usually large flow curvature and wall shear near the cylinder wall, obvious compression flow in the upstream of the cylinder, and stretching flow, flow separation and vortex shedding in the wake flow field. Due to the complexity of the flow, materials with different rheological parameters, geometric sizes and shapes, etc., it exhibits complex flow phenomena, such as upstream recirculation (Kenney *et al.*, 2013; Shi *et al.* 2015; Zhao *et al.* 2016; Qin *et al.* 2019; Haward *et al.* 2021, 2022), flow asymmetry (Nolan *et al.* 2016; Haward *et al.* 2018, 2020), etc. The related simulation about viscoelatic flow over a blunt body is complicated. Following, we review this flow simulation at moderate Reynolds numbers (*Re*) according to the timeline.

Oliveira (2001) numerically simulated the viscoelastic flow over a cylinder at *Re* from 50 to 120 for the first time. They adopted the viscoelastic constitutive model of FENE-MCR. The FENE-MCR model is simplified on the basis of the FENE-P model, which makes it more stable in simulation. It is not necessary for the author to add any artificial dissipation in the transport equations of conformation tensor or elastic stress tensor. For $L$ = 10, the maximum calculated *Wi* of the author is as high as 80. The author has a good understanding of the characteristics of polymer addition to suppress flow instability and the characteristics of inhibiting flow fluctuation frequency. For small molecular chain ($L$=10), the addition of polymer could reduce drag of the cylinder. However, when the molecular chain is long ($L$ = 30), polymer addition increases drag of the cylinder.

Sahin & Owens (2004) focuses on the first transition of flow (the beginning of vortex shedding) affected by viscoelasticity. Sahin & Owens (2004) used the linear stability analysis method to study effect of fluid's viscoelasticity on flow stability. They found that the addition of polymer could increase the transitional Reynolds number ($Re_c$). Sahin & Owens (2004) also adopted the constitutive simulation of FENE-MCR, without introducing artificial dissipation in the transport equations of conformation tensor or elastic stress tensor.

Later, Richter *et al.* (2010) used the FENE-P model to simulate viscoelastic flow over a circular cylinder in two-dimensions (*Re* = 100) and three-dimensions (*Re* = 300). However, due to the singularity of FENE-P model, the numerical calculation is unstable, and they directly introduce artificial dissipation into the conformation tensor equations. The magnitude of artificial dissipation satisfies a dimensionless number, Schmidt number $Sc$>10. Simulation of Richter *et al.* (2010) also found that adding polymer can inhibit the flow instability and weaken the frequency of flow fluctuation, which is similar to the results of Oliveira (2001). When the molecular chain is small ($L$ = 10), the addition of polymer could reduce the drag of cylinder. However, when the molecular chain is long ($L$ = 100), the addition of polymer could increase drag of the cylinder, and the drag coefficient is as high as 2.7 at *Wi* = 10, while that in Newtonian fluid is about 1.34. Through three-dimensional numerical simulation of *Re* = 300, Ritchter *et al.* (2010) found that adding polymer could inhibit the instability of three-dimensional flow. The flow in Newtonian fluid is regarded as Mode B instability. The wave length in spanwise direction is about 0.9$D$ ($D$ is diameter of cylinder). After the addition of low molecular weight ($L$ = 10) polymer, the Mode B type instability changes to the Mode A type instability (the spanwise wavelength becomes longer). However, when the polymer with high molecular weight ($L$ = 100) is added, the spanwise fluctuation is greatly suppressed, while the wavelength becomes shorter. The three-dimensional behaviour obtained by direct numerical simulation coincides well with three-dimensional linear stability analysis (Ritchter *et al.* 2012a). Ritchter *et al.* (2012b) numerically simulated viscoelastic flow over a three-dimensional cylinder with *Re* of 3900. At this *Re*, the wake field in Newtonian fluid is turbulence. The addition of polymer with molecular chain length of $L$ = 10 has weak inhibition effect on turbulence.. But after the addition of polymer with molecular chain length of $L$ = 100, the turbulence is obviously suppressed, and even the turbulence state is transformed into Mode A instability. As the turbulence in wake field is suppressed, the pressure at the rear of the cylinder increases obviously. For $L$ = 10, the drag of cylinder decreases slightly. However, for $L$ = 50 and $L$ = 100, the drag of cylinder increases compared with that of Newtonian fluid. In this numerical simulation, the authors introduced artificial dissipation in the transport equations of conformation tensor, and the Schmitt number (*Sc*) corresponding to artificial dissipation is 0.69.

Xiong' research group has done a lot of related numerical simulation research work in this field. Xiong *et al.*(2010, 2011, 2013) simulated the flow of Oldroyd-B fluid around a cylinder at a wide *Re* range of 0.01 to 50,000. Their simulation are carried out in two-dimensional space. Compared with Newtonian fluid, the maximum drag reduction rate is about 50%, and the corresponding *Re* is about 2,000. The addition of polymer inhibits the flow instability or two-dimensional turbulence at low *Wi*. However, it makes the flow unstable again when the *Wi* is high. In these studies, the cylindrical wall is treated by immersed boundary method using the finite difference method, while the elastic stress at the wall is set to zero directly. To makes the numerical calculation stable, artificial dissipation is introduced into the conformational tensor transport equation. Late, they used finite volume method for other study (Xiong *et al.* 2018, 2019; Peng *et al.* 2020, 2022). Xiong *et al.* (2018) studied the effect of polymer addition on hydrofoil with attack angle. Numerical simulation shows that polymer addition could restrain the flow asymmetry and reduce the lift of hydrofoil. Xiong *et al.* (2019) studied how polymer addition affects vortex-induced vibration of cylinders. This problem is a simple fluid-solid coupling problem. Peng *et al.* (2020) studied the influence of polymer addition on the flow around side-by-side cylinders. In these simulations, artificial dissipation are introduced, and the Schmidt number of artificial dissipation is set as 10.

In previous studies, artificial dissipation was often introduced directly because of the instability of viscoelastic flow calculation. However, its impact on the flow calculation simulation has not been rigorously evaluated. Evaluation of it is helpful to reconfirm the previous research work. At extremely low *Re*, some stabilization techniques are often adopted, such as logarithmic reconstruction method, the square root tensor method and so on (Fattal & Kupferman 2005; Balci *et al.* 2011; Afonso *et al.* 2012). However, numerical tests indicate that various stabilization methods behave differently at high *Wi*. Recently, the logarithmic reconstruction method has been introduced into the calculation of viscoelastic fluid flow around a cylinder with *Re* of 100 (Peng *et al.* 2021). Flow around a cylinder of Giesekus (including Oldroyd-B) fluid. In their numerical simulation, the logarithmic reconstruction method (Log) was introduced into the calculation. No artificial dissipation is required. This method has been used to simulate polar Reynolds numbers in the past. Giesekus model is a constitutive model which shows both viscoelastic and shear-thinning. Numerical simulation shows that the introduction of fluid viscoelasticity can suppress the flow instability. However, under the condition of high elasticity, the flow will become unstable again, which was found in previous experiments (Nolan *et al.* 2016).

The introduction of logarithmic reconstruction method enables us to calculate on a high *Wi*. However, when it is introduced into the calculation at *Re*, this stabilization method still needs to be strictly evaluated.

The two classical stabilization techniques are the logarithmic reconstruction method (*Log*) and the square root reconstruction method (*Sqrt*).In this paper, the flow of FENE-P viscoelastic fluid around a cylinder with Reynolds number of 100 is simulated numerically. In this study, three stabilization techniques are considered, namely, traditional method plus artificial dissipation (*Td*), Log reconstruction method (*Log*) and the square root tensor method (*Sqrt*).

## 2. Problem formulation

### 2.1 Governing equations

The dimensionless governing equations for incompressible fluids with polymer additives could be expressed as follows:

$$\nabla \cdot \mathbf{u} = 0, \quad (1)$$

$$\frac{\partial \mathbf{u}}{\partial t} + \mathbf{u} \cdot \nabla \mathbf{u} = -\nabla p + \frac{\beta}{Re} \Delta \mathbf{u} + \frac{1-\beta}{Re \cdot Wi} \nabla \cdot \boldsymbol{\tau}^{p}. \quad (2)$$

This expression is very similar to the dimensionless Navier-Stokes equation for incompressible Newtonian flow, except for polymer stress. is a term that represents additional body stress due to the elasticity of the polymer in the flow. To describe the degree of viscoelasticity, the Weissenberg number (*Wi*) is defined as the ratio of the characteristic polymer relaxation time scale to the characteristic flow time scale, where for this case it is the cylinder diameter. Reynolds number (*Re*) is the ratio of inertial force to viscous force, $\beta$ refers to the ratio of the zero shear rate viscosity of the solvent to the zero shear rate viscosity of the total solution ($v_s$ is the viscosity contribution from the solvent, and $v_p$ is the viscosity contribution from the polymer).

In order to close the equations, a polymer stress model must be introduced, and for this work, a molecule-based FENE-P model is used. The model is similar to a single member of a polymer of diluted concentration, as a single dumbbell connected to a finitely stretchable nonlinear elastic spring, and through the balance of the forces acting on the beads, the kinetic theory can be used to determine the polymer stress $\boldsymbol{\tau}^{p}$ expressed as (see Bird, Armstrong & Hassager 1987):

$$\boldsymbol{\tau}^{p} = \frac{\mathbf{c}}{1-\frac{tr(\mathbf{c})}{L^2}} - \frac{\mathbf{I}}{1-\frac{3}{L^2}}. \quad (3)$$

In this equation, $L$ refers to the maximum polymer extensibility, which is non- dimensionalized by the equilibrium length of a linear spring ( $(kT / H)^{1/2}$ ) where $T$ is the absolute temperature, $k$ is Boltzmann's constant and $H$ is the Hookean spring constant for an entropic spring. Also, $\mathbf{c}$ represents the averaged polymer conformation tensor (also scaled by the equilibrium Hookean spring length), which is defined as the preaveraged diadic product of the polymer end-to-end vector and is governed by the hyperbolic transport equation shown below (for discussion of the mathematical character of the FENE-P fluid equation, see Purnode & Legat 1996).

$$\frac{\partial \mathbf{c}}{\partial t}+\mathbf{u}\cdot\nabla\mathbf{c}-\mathbf{c}\cdot\nabla\mathbf{u}-\nabla\mathbf{u}^T\cdot\mathbf{c}=-\frac{\boldsymbol{\tau}^p}{Wi}. \tag{4}$$

The FENE-P model is chosen for this work based on its ability to properly represent the finite extensibility, and thus the bounded stress, of the polymers. For problems with large $Wi$ and large strain rates, this feature is required in order to obtain bound solutions, and linear springs such as the Oldroyd-B constitutive model cannot be faithfully used. Moreover, the FENE-P model has been used in many previous studies involving high-Reynolds-number viscoelastic flows, and its ability to provide accurate physical insight into these types of problems has been demonstrated (see Azaiez & Homsy 1994b; Sureshkumar et al. 1997; Kumar & Homsy 1999; Dubief et al. 2004, 2005; Dimitropoulos et al. 2005, 2006).

## 2.2 Numerical methods

### A. Implicit calculation of $C_{kk}$

The governing equations are solved by the open-source CFD platform OpenFOAM (Weller *et al.*, 1998) and the rheotool toolbox (Pimenta & Alves 2018). In order to ensure the boundedness of $c_{kk}$, an implicit algorithm is used for pre-calculation before each time step (Ritchter *et al.*, 2010). Tracing the transport equation of the conformation tensor yields,

$$\frac{\partial c_{kk}}{\partial t}+(\mathbf{u}\cdot\nabla)c_{kk}=tr\left[(\nabla\mathbf{u})\cdot\mathbf{c}+\mathbf{c}\cdot(\nabla\mathbf{u})^T\right]-\frac{(c_{kk}-3)L^2}{\lambda(L^2-c_{kk})}. \tag{5}$$

By defining

$$\varphi=-\ln\left(1-\frac{c_{kk}}{L^2}\right), \tag{6}$$

Eq. (5) can be rewritten as,

$$\frac{\partial\varphi}{\partial t}+u_j\frac{\partial\varphi}{\partial x_j}=\frac{e^\varphi}{L^2}\left(c_{kj}\frac{\partial u_k}{\partial x_j}+c_{jk}\frac{\partial u_j}{\partial x_k}\right)+\frac{e^\varphi}{\lambda L^2}\left(3+L^2-L^2e^\varphi\right). \tag{7}$$

The scalar $\varphi$ is solved before each time step, and then saved for the calculation of the next time step.

Three methods are used to solve the conformation tensor transport equations in this paper, traditional method, logarithmic reconstruction method, and square reconstruction method. Following, we display these formulas below, respectively.

### B. Traditional method

The conformation tensor transport equations are solved directly. For the stability of numerical calculation, we introduce artificial dissipation to the right of the conformation tensors, as follows,

$$\frac{\partial \mathbf{c}}{\partial t}+\mathbf{u}\cdot\nabla\mathbf{c}-\mathbf{c}\cdot\nabla\mathbf{u}-\nabla\mathbf{u}^T\cdot\mathbf{c}=-\frac{\boldsymbol{\tau}^p}{Wi}+\kappa\Delta\mathbf{c}. \tag{8}$$

### C. Logarithmic reconstruction method

The log-conformation tensor approach consists in a change of variable when evolving in time the polymeric extra-stress and it was devised to tackle the numerical instability faced at high Weissenberg number flows. In the log-conformation tensor methodology, a new tensor ( $\mathbf{\Theta}$ ) is defined as the natural logarithm of the conformation tensor

$$\mathbf{\Theta} = \ln(\mathbf{c}) = \mathbf{R}\ln(\mathbf{\Lambda})\mathbf{R}^T. \tag{9}$$

In Eq. (5), the conformation tensor was diagonalized ( $\mathbf{c} = \mathbf{R}\mathbf{\Lambda}\mathbf{R}^T$ ) because it is positive definite, where $\mathbf{R}$ is a matrix containing in its columns the eigenvectors of $\mathbf{c}$ and $\mathbf{\Lambda}$ is a matrix whose diagonal elements are the respective eigenvalues resulting from the decomposition of $\mathbf{c}$. Eq. (9) written in terms of ( $\mathbf{\Theta}$ ) becomes

$$\frac{\partial \mathbf{\Theta}}{\partial t} + \mathbf{u}\cdot\nabla\mathbf{\Theta} = \Omega\mathbf{\Theta} - \mathbf{\Theta}\Omega + 2\mathbf{B} + \frac{1}{Wi}g(\mathbf{\Theta}), \tag{10}$$

where g( $\mathbf{\Theta}$ ) is a model-specific tensorial function depending on $\mathbf{\Theta}$ and

$$\mathbf{B} = \mathbf{R}\begin{bmatrix} m_{xx} & 0 & 0 \\ 0 & m_{yy} & 0 \\ 0 & 0 & m_{zz} \end{bmatrix}\mathbf{R}^T \tag{11}$$

$$\mathbf{\Omega} = \mathbf{R}\begin{bmatrix} 0 & \omega_{xy} & \omega_{xz} \\ -\omega_{xy} & 0 & \omega_{yz} \\ -\omega_{xz} & -\omega_{yz} & 0 \end{bmatrix}\mathbf{R}^T \tag{12}$$

$$\mathbf{M} = \mathbf{R}\nabla\mathbf{u}^T\mathbf{R}^T = \begin{bmatrix} m_{xx} & m_{xy} & m_{xz} \\ m_{yx} & m_{yy} & m_{yz} \\ m_{xz} & m_{zy} & m_{zz} \end{bmatrix} \tag{13}$$

$$\omega_{\mathrm{ij}} = \frac{\Lambda_j m_{ij} + \Lambda_i m_{ji}}{\Lambda_j - \Lambda_i} \tag{14}$$

Especially to FENE-P model,

$$g(\mathbf{\Theta}) = \frac{L^2}{L^2-3}e^{-\mathbf{\Theta}} - \frac{L^2}{L^2 - tr(e^{\mathbf{\Theta}})}\mathbf{I}. \tag{15}$$

**D. Square reconstruction method**

In order to ensure the stability of numerical calculation, the square root reconstruction method is adopted. A new symmetric tensor **b** is introduced into by $\mathbf{c}=\mathbf{b}\cdot\mathbf{b}^T$, which satisfies,

$$\frac{\partial \mathbf{b}}{\partial t} + (\mathbf{u}\cdot\nabla)\mathbf{b} = \mathbf{b}\nabla\mathbf{u} + \mathbf{ab} + \frac{1}{2\lambda}\left(\theta/\mathbf{b} - \mathbf{b}e^{\varphi}\right). \tag{16}$$

Note that **a** in Eq. (16) is an anti-symmetric tensor, which could be written in the form of components as,

$$\mathbf{a} = \begin{pmatrix} 0 & a_{12} & a_{13} \\ -a_{12} & 0 & a_{23} \\ -a_{13} & -a_{23} & 0 \end{pmatrix}. \tag{17}$$

The components of **a** could be calculated by solving the following equations,

$$\begin{aligned} &\left(b_{11}+b_{22}\right)a_{12}+b_{23}a_{13}-b_{31}a_{23}=w_1, \\ &b_{23}a_{12}+\left(b_{11}+b_{33}\right)a_{13}+b_{12}a_{23}=w_2, \\ &-b_{13}a_{12}+b_{12}a_{13}+\left(b_{22}+b_{33}\right)a_{23}=w_3, \end{aligned} \tag{18}$$

where

$$\begin{aligned} w_1 &= \left(b_{12}u_{1,1}-b_{11}u_{2,1}\right)+\left(b_{22}u_{1,2}-b_{12}u_{2,2}\right)+\left(b_{23}u_{1,3}-b_{13}u_{2,3}\right), \\ w_2 &= \left(b_{13}u_{1,1}-b_{11}u_{3,1}\right)+\left(b_{33}u_{1,3}-b_{13}u_{3,3}\right)+\left(b_{23}u_{1,2}-b_{12}u_{3,2}\right), \\ w_3 &= \left(b_{13}u_{2,1}-b_{12}u_{3,1}\right)+\left(b_{23}u_{2,2}-b_{22}u_{3,2}\right)+\left(b_{33}u_{2,3}-b_{23}u_{3,3}\right). \end{aligned} \tag{19}$$

where $u_{i,j}$ is the components of $\nabla\mathbf{u}$. For a detailed description of this method, the reader can refer to Balci *et al.* (2011).

E. **Solution method of pressure-velocity coupling**

In the OpenFOAM toolbox, common algorithms for pressure-velocity coupling are SIMPLE and SIMPLEC for steady-state solvers and either PISO or PIMPLE (a combination of SIMPLE(C) and PISO) for transient solvers. From the benchmark cases performed in Ref. [2], it was observed that SIMPLEC was particularly suitable for transient viscoelastic fluid flows at low Reynolds numbers, regarding stability and accuracy. The continuity equation, implicit in the pressure variable, derived for SIMPLEC (a more detailed derivation is presented in Ref. [2]) leads to

$$\nabla\cdot\left(\frac{1}{a_P-H_1}\left(\nabla p\right)_P\right)=\nabla\cdot\left[\frac{\mathbf{H}}{a_P}+\left(\frac{1}{a_P-H_1}-\frac{1}{a_P}\right)\left(\nabla p^*\right)_P\right] \tag{20}$$

where $a_P$ are the diagonal coefficients from the momentum equation, $H_1=-\sum_{nb}a_{nb}$ is an operator representing the negative sum of the off-diagonal coefficients from momentum equation, $\mathbf{H}=-\sum_{nb}a_{nb}\mathbf{u}^*_{nb}+\mathbf{b}$ is an operator containing the offdiagonal contributions, plus source terms (except the pressure gradient) of the momentum equation and $p^*$ is the pressure field known from the previous timestep or iteration. Accordingly, the equation to correct the velocity after obtaining the continuity-compliant pressure field from Eq. (3.15) is

$$\mathbf{u}=\frac{\mathbf{H}}{a_P}+\left(\frac{1}{a_P-H_1}-\frac{1}{a_P}\right)\left(\nabla p^*\right)_P-\frac{1}{a_P-H_1}\left(\nabla p\right)_P. \tag{21}$$

The SIMPLEC algorithm has better calculation accuracy and stability for the unsteady calculation of low Reynolds number viscoelastic fluids. However, the calculation accuracy of this method will reduce the accuracy of the non-steady calculation under the medium Reynolds number; the corresponding PISO algorithm has higher accuracy, but the stability of the viscoelastic fluid calculation will be reduced. The PIMPLE algorithm is a pressure correction algorithm with the SIMPLE algorithm embedded in the outer layer of the PISO algorithm. This method has both the stability of the SIMPLE algorithm and the high calculation accuracy of the PISO algorithm.

Importantly, in order to avoid the onset of checkerboard fields, the pressure gradient terms involved in the computation of face velocities, i.e., in Eqs. (3.15) and (3.16), are directly evaluated using the pressure on the cells straddling the face, in a Rhie-Chow-like procedure (more details in Ref. [2]). Nonetheless, when Eq. (3.16) is used to correct the cell-centered velocity field, the pressure gradient terms are computed ”in the usual way”, for example using Green-Gauss integration.

Rhie-Chow methods used to avoid checkerboard fields, as the one described in the previous paragraph, are known to be affected by the use of small time-steps and they also present time-step dependency on steady-state results [11]. In OpenFOAM solvers, a common strategy to avoid such effects is to add a corrective term to face-interpolated velocities, through functions ddtPhiCorr() or ddtCorr(). Recently, in foam-extend the time-step dependency was solved in a different way, by removing the transient term contribution from the aP coefficients of the momentum equation [12]. However, this approach may be problematic when used with the SIMPLEC algorithm, since a division by zero is prone to happen. In rheoTool, we keep using the added corrective term, although, as mentioned in Ref. [2], this term can be improved in order to more efficiently avoid the small time-step dependency of steady-state solutions.

## F. Problem specification and boundary conditions

The problem definition is that of viscoelastic flow around a circular cylinder at $Re$ = 100. At present, simulations are limited to this moderate-Reynolds-number range due to the large amount of computation time required to probe higher $Re$. In general, cylinder flow is rich in physical effects such as shear layers, recirculation regions, boundary layers and vortex dynamics, thus making this problem ideal for studying complex viscoelastic effects. Furthermore, as the Reynolds number is increased, we know that for a Newtonian fluid the flow type changes dramatically, starting from steady laminar flow, changing to unsteady two-dimensional vortex shedding, then going through several stages of three-dimensional transition before finally reaching full turbulence (see Williamson 1996b). As a result, these different stages also present opportunities to investigate the effect of viscoelasticity under many different circumstances. Because the Newtonian counterpart has been studied extensively in the past (much of which is reviewed in Williamson 1996b), comparisons between Newtonian and non-Newtonian flows can be easily made.

For the cases chosen, Newtonian flow at $Re$ = 100 lies within the two-dimensional laminar vortex shedding regime. For each different case, a slightly different mesh was used to perform the calculations. A schematic of the $x$-$y$ plane of the respective domains, denoted by Mesh 1 and Mesh 2, is shown in figure 1. The primary difference between the domains is that for Mesh 2, the downstream exit boundary was extended from 16.5$D$ in Mesh 1 to 50$D$. Because the fluid motions are purely two-dimensional at a Reynolds number of 100, the spanwise domain length is set at 1$D$ and is discretized using only one cell for all $Re$ =100 cases.

## G. Some formula definitions

In the present study, the length and velocity are normalized by $D$ and $u_{in}$, respectively. Time,

pressure, stresses, and vorticity are scaled by $D/u_{in}$, $\rho u_{in}^2$, $\rho u_{in}^2$, $u_{in}/D$, respectively. A group of dimensionless parameters is adopted, including Wi, Re, β , St, Cd and Cl. Wi represents the ratio of elastic to viscous forces [40], while Re represents the ratio between inertia and viscous forces. St denotes the dimensionless frequency of vortex shedding. β is the viscosity ratio between the solvent and the solution at zero shear rate, a measurement of polymer con-centration and molecular characteristics of polymer. Cd and Cl are the lift and drag force coefficients acted on the cylinder, respectively. These dimensionless parameters are defined as follows:

Wi = λ uinD ,

Re = ρ uinD η 0,

β = η p η 0,

St = fsDuin,

Cd = 2Fx ρ U2inD,

Cl = 2Fy ρ U2inD

where fs is the shedding frequency, and η 0= η p+ η s is the summation of the polymer and solvent viscosities at the zero-shear rate, Fx and Fy are the drag and lift of the fluid acted to the cylinder. The drag force consists of three parts, i.e., the pressure drag force Fpressure, the viscous drag force Fviscous, and the elastic drag force Fpolymer, which corresponding to the pressure drag coefficient Cpressured , the viscous drag coefficient Cviscousd , the elastic drag coefficient Cpolymerd . The drag components can be calculated by integrating the corresponding stress component along the cylinder surface, i.e., Fpressure =。

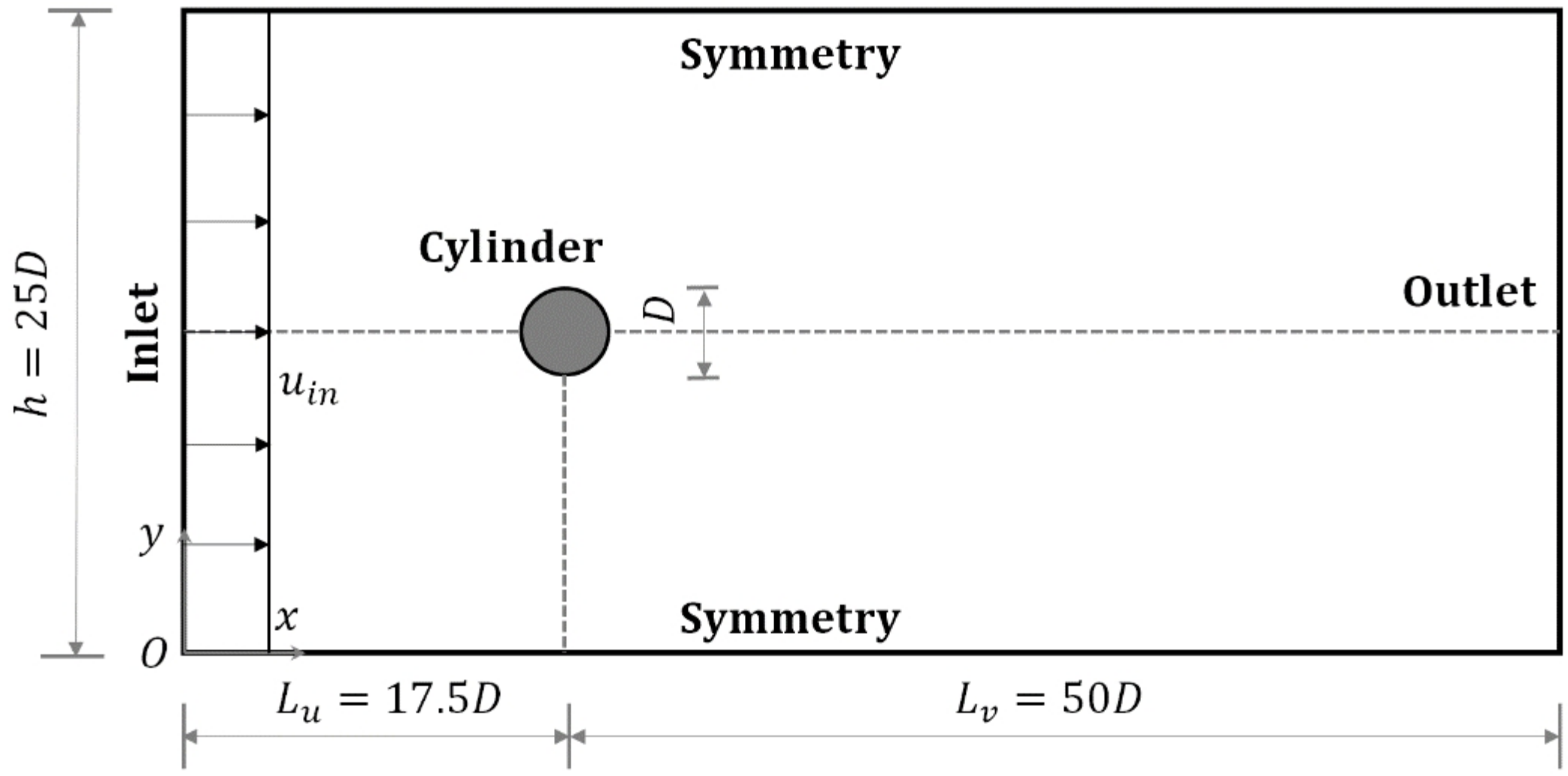

**Fig.1** The calculated spatial domain.

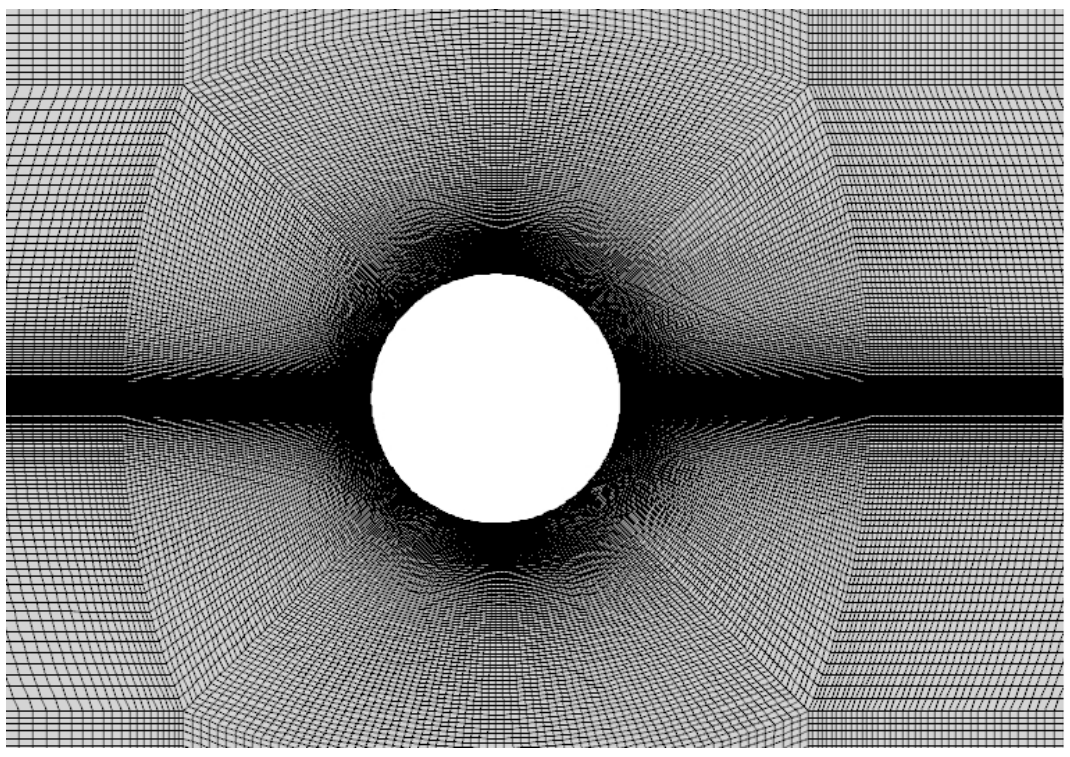

**Fig.2** Mesh distribution near the cylinder wall.

## 3. Results and discussions

### 3.1 **Small chain length for three different methods**

The calculation results of molecular chain $L$ = 10 is often used to check the accuracy of viscoleatic flow algorithm. Previous numerical tests of the FENE-MCR model is table, then no artificial diffusion is needed to add. However, the FENE-P model simulation test indicates it is not stable while needs add artificial diffusion. Previous numerical simulations indicate the drag is reduced compared with that in Newtonian fluid at the same fixed Reynolds number, while suppress flow instability for small polymer length viscoelastic fluids. The results with tranditional method for Sc = 10 are list in table 1, then compared with previous publications of Richter *et al.* (2010) and Xiong *et al.* (2019), list in table 2. Our simulation results coincide with previous simulation results. The drag is reduced with that in Newtonian fluid. For example, the average drag coefficient of Wi = 80 is 1.212, while it is 1.361 in Newtonian fluid. Richter *et al.* and Xiong *et al.* are obtained $\overline{C_d}$ are 1.216 and 1.227, respectively. The reduction of drag comes from two aspects, one is the reduction of pressure drag ( $\overline{C_{dp}}$ ), the other is the reduction of wall friction drag ( $\overline{C_{dv}}$ ). Small-amplitude polymer stress drag occurs ( $\overline{C_{de}}$ ). The flow instability is suppressed. The root mean square of the lift coefficient ( $C_{lrms}$ )could reflect the fluctuating intensity of the flow field at the cylinder wall. The velocity at 1D after the tail end point of the cylinder is monitored and counted, which can partly reflect the fluctuation intensity of wake field. $u_{rms}$ and $v_{rms}$ are list in table 1 and table 2. $C_{lrms}$ , $u_{rms}$ and $v_{rms}$ decrease as *Wi* increase. Strouhal number(*St*) is a dimensionless number of fluctuation frequency of wake vortex shedding. *St* increase then decrease as *Wi* increase. The results of square root reconstruction method (Sqrt) are list in table 3. The results of Sqrt are simular to Tm. The results of logarithmic reconstruction method (Log) are list in table 4. The results of Log are similar to Tm or Sqrt.

Table 1. Results for increasing *Wi* of *Re* = 100, *L* = 10 and *Sc* = 10 for *Td* method.

| $Wi$ | $L$ | $\overline{C_{dp}}$ | $\overline{C_{dv}}$ | $\overline{C_{de}}$ | $\overline{C_d}$ | $C_{lrms}$ | $St$ | $u_{rms}$ | $v_{rms}$ |
|---|---|---|---|---|---|---|---|---|---|
| 0 | 10 | 1.014 | 0.347 | 0 | 1.361 | 0.235 | 0.1669 | 0.0458 | 0.3091 |
| 0.5 | 10 | 1.027 | 0.311 | 0.027 | 1.365 | 0.224 | 0.1665 | 0.0473 | 0.3044 |
| 1 | 10 | 1.027 | 0.304 | 0.025 | 1.356 | 0.202 | 0.1660 | 0.0406 | 0.2795 |
| 2 | 10 | 1.011 | 0.295 | 0.023 | 1.329 | 0.155 | 0.1637 | 0.0279 | 0.2272 |
| 4 | 10 | 0.982 | 0.286 | 0.023 | 1.291 | 0.096 | 0.1614 | 0.0161 | 0.1640 |
| 6 | 10 | 0.966 | 0.282 | 0.023 | 1.271 | 0.074 | 0.1604 | 0.0120 | 0.1327 |
| 10 | 10 | 0.948 | 0.278 | 0.023 | 1.249 | 0.052 | 0.1601 | 0.0079 | 0.1038 |
| 20 | 10 | 0.930 | 0.276 | 0.023 | 1.229 | 0.034 | 0.1610 | 0.0049 | 0.0812 |
| 40 | 10 | 0.919 | 0.276 | 0.023 | 1.218 | 0.026 | 0.1629 | 0.0039 | 0.0708 |
| 80 | 10 | 0.913 | 0.276 | 0.023 | 1.212 | 0.022 | 0.1652 | 0.0036 | 0.0663 |

Table 2. Results compared with Richter *et al.*(2010) and Xiong *et al.* (2019) for increasing *Wi* of *Re* = 100, *L* = 10 and *Sc* = 10 for *Td* method.

| *Wi* | $\overline{C_d}$ | | | *St* | | | $u_{rms}$ | | $v_{rms}$ | |
|---|---|---|---|---|---|---|---|---|---|---|
| | Present | Ref.1 | Ref.2 | Present | Ref.1 | Ref.2 | Present | Ref.1 | Present | Ref.1 |
| 0 | 1.361 | 1.343 | 1.361 | 0.1669 | 0.1677 | 0.1669 | 0.0458 | 0.0498 | 0.3091 | 0.3169 |
| 0.5 | 1.365 | 1.351 | 1.362 | 0.1665 | 0.1657 | 0.1658 | 0.0473 | 0.0480 | 0.3044 | 0.3009 |
| 1 | 1.356 | 1.346 | 1.363 | 0.1660 | 0.1641 | 0.1645 | 0.0406 | 0.0401 | 0.2795 | 0.2745 |
| 2 | 1.329 | - | 1.34 | 0.1637 | - | 0.1625 | 0.0279 | - | 0.2272 | - |
| 4 | 1.291 | - | 1.304 | 0.1614 | - | 0.1601 | 0.0161 | - | 0.1640 | - |
| 6 | 1.271 | - | - | 0.1604 | - | - | 0.0120 | - | 0.1327 | - |
| 10 | 1.249 | 1.25 | 1.265 | 0.1601 | 0.1603 | 0.1582 | 0.0079 | 0.0079 | 0.1038 | 0.1015 |
| 20 | 1.229 | 1.233 | 1.246 | 0.1610 | 0.1585 | 0.1584 | 0.0049 | 0.0043 | 0.0812 | 0.0762 |
| 40 | 1.218 | 1.222 | 1.233 | 0.1629 | 0.1569 | 0.1592 | 0.0039 | 0.0028 | 0.0708 | 0.0606 |
| 80 | 1.212 | 1.216 | 1.227 | 0.1652 | 0.1566 | 0.1596 | 0.0036 | 0.0022 | 0.0663 | 0.0524 |

Table 3. Results for increasing *Wi* of *Re* = 100, *L* =10 for Sqrt method.

| *Wi* | *L* | $\overline{C_{dp}}$ | $\overline{C_{dv}}$ | $\overline{C_{de}}$ | $\overline{C_d}$ | $C_{lrms}$ | *St* | $u_{rms}$ | $v_{rms}$ |
|---|---|---|---|---|---|---|---|---|---|
| 0 | 10 | 1.014 | 0.347 | 0 | 1.361 | 0.235 | 0.1669 | 0.0458 | 0.3091 |
| 0.5 | 10 | 1.021 | 0.317 | 0.020 | 1.358 | 0.218 | 0.1673 | 0.0461 | 0.3004 |
| 1 | 10 | 1.016 | 0.312 | 0.015 | 1.343 | 0.195 | 0.1663 | 0.0394 | 0.2750 |
| 2 | 10 | 0.999 | 0.306 | 0.01 | 1.315 | 0.155 | 0.1644 | 0.0282 | 0.2276 |
| 4 | 10 | 0.974 | 0.300 | 0.007 | 1.281 | 0.109 | 0.1622 | 0.0184 | 0.1719 |
| 6 | 10 | 0.960 | 0.297 | 0.006 | 1.263 | 0.085 | 0.1612 | 0.0144 | 0.1431 |
| 10 | 10 | 0.945 | 0.294 | 0.004 | 1.243 | 0.062 | 0.1606 | 0.0093 | 0.1165 |
| 20 | 10 | 0.931 | 0.288 | 0.002 | 1.221 | 0.040 | 0.1597 | 0.0041 | 0.0846 |
| 40 | 10 | 0.924 | 0.284 | 0.001 | 1.210 | 0.026 | 0.1618 | 0.0029 | 0.0669 |
| 80 | 10 | 0.922 | 0.283 | 0 | 1.205 | 0.021 | 0.1634 | 0.0027 | 0.0644 |

Table 4. Results for increasing *Wi* of *Re* = 100, *L* = 10 for Log method.

| $Wi$ | $L$ | $\overline{C_{dp}}$ | $\overline{C_{dv}}$ | $\overline{C_{de}}$ | $\overline{C_d}$ | $C_{lrms}$ | $St$ | $u_{rms}$ | $v_{rms}$ |
|---|---|---|---|---|---|---|---|---|---|
| 0 | 10 | 1.014 | 0.347 | 0 | 1.361 | 0.235 | 0.1669 | 0.0458 | 0.3091 |
| 0.5 | 10 | 1.022 | 0.314 | 0.022 | 1.358 | 0.216 | 0.1672 | 0.0437 | 0.2986 |
| 1 | 10 | 1.019 | 0.309 | 0.017 | 1.345 | 0.196 | 0.1664 | 0.0386 | 0.2678 |
| 2 | 10 | 1.002 | 0.302 | 0.013 | 1.317 | 0.166 | 0.1643 | 0.0277 | 0.2257 |
| 4 | 10 | 0.976 | 0.296 | 0.012 | 1.284 | 0.119 | 0.1621 | 0.0180 | 0.1688 |
| 6 | 10 | 0.964 | 0.294 | 0.007 | 1.265 | 0.093 | 0.1605 | 0.0135 | 0.1397 |
| 10 | 10 | 0.947 | 0.291 | 0.003 | 1.240 | 0.059 | 0.1603 | 0.0082 | 0.1127 |
| 20 | 10 | 0.929 | 0.285 | 0.002 | 1.216 | 0.035 | 0.1599 | 0.0035 | 0.0792 |
| 40 | 10 | 0.922 | 0.283 | 0.001 | 1.206 | 0.026 | 0.1623 | 0.0026 | 0.0697 |
| 80 | 10 | 0.920 | 0.282 | 0 | 1.202 | 0.024 | 0.1635 | 0.0023 | 0.0612 |

Table 5. Summary of three stabilization techniques for small chain length

| $Wi$ | $L$ | $\overline{C_d}$ | | | $C_{lrms}$ | | |
|---|---|---|---|---|---|---|---|
| Stabilization techniques | | $Td$ | $Sqrt$ | $Log$ | $Td$ | $Sqrt$ | $Log$ |
| 0 | | 1.361 | 1.361 | 1.361 | 0.235 | 0.235 | 0.235 |
| 0.5 | 10 | 1.365 | 1.358 | 1.358 | 0.224 | 0.218 | 0.216 |
| 1 | 10 | 1.356 | 1.343 | 1.358 | 0.202 | 0.195 | 0.196 |
| 2 | 10 | 1.329 | 1.315 | 1.345 | 0.155 | 0.155 | 0.166 |
| 4 | 10 | 1.291 | 1.281 | 1.317 | 0.096 | 0.109 | 0.119 |
| 6 | 10 | 1.271 | 1.263 | 1.284 | 0.074 | 0.085 | 0.093 |
| 10 | 10 | 1.249 | 1.243 | 1.265 | 0.052 | 0.062 | 0.059 |
| 20 | 10 | 1.229 | 1.221 | 1.240 | 0.034 | 0.040 | 0.035 |
| 40 | 10 | 1.218 | 1.210 | 1.216 | 0.026 | 0.026 | 0.026 |
| 80 | 10 | 1.212 | 1.205 | 1.206 | 0.022 | 0.021 | 0.024 |

When $L$ is very small in the FENE-P model, there exits small difference in the simulation results of the three stabilization technologies. The addition of artificial dissipation has little effect on the simulation results of viscoelastic flow with small molecular chain length, as list in table 5.

### 3.2 Long chain length

A. **Traditional method**

$L$ = 100 is often considered as a long chain length polymer solution in the studies of Xiong's group and Rithter's study. In previous, the traditional method is used while $Sc$ is set as 10. The cylinder's drag coefficient ($\overline{C_d}$) is 2.738 for ($Re$, $Wi$, $L$, $Sc$) = (100, 10, 100, 10) in our this study, compared that Richter et al. (2010) get it as 2.7. The results for another $Wi$ at this $Sc$ are list in table 6. The drag increase main results from the pressure difference drag ($\overline{C_{dp}}$) increase. For that $Sc$, $\overline{C_{dv}}$ decrease greatly. However, $\overline{C_{de}}$ increase greatly. The flow pulsation on the cylinder wall is obviously weakened. At $Wi$ = 6, $C_{lrms}$ is as low as 0, while $u_{rms}$ is also 0 and $v_{rms}$ is only 0.0012. The fluctuation frequency of flow field is obviously weakened when $Wi$ is no more than 6.. At $Wi$ = 6, the strouhal number is only 0.0883,which is about half that in Newtonian fluid. The strouhal number for $Wi$ = 10 is a little higher than that for $Wi$ = 6.

Table 6. Results for increasing $Wi$ of $Re$ =100, $L$ = 100, $Sc$ = 10 for Td.

| $Wi$ | $L$ | $Sc$ | $\overline{C_{dp}}$ | $\overline{C_{dv}}$ | $\overline{C_{de}}$ | $\overline{C_d}$ | $C_{lrms}$ | $St$ | $u_{rms}$ | $v_{rms}$ |
|---|---|---|---|---|---|---|---|---|---|---|
| 0 | 100 | 10 | 1.014 | 0.347 | 0 | 1.361 | 0.235 | 0.1669 | 0.0458 | 0.3091 |
| 0.5 | 100 | 10 | 1.113 | 0.254 | 0.071 | 1.438 | 0.244 | 0.1583 | 0.0525 | 0.3184 |
| 1 | 100 | 10 | 1.422 | 0.13 | 0.147 | 1.699 | 0.195 | 0.1431 | 0.0417 | 0.2588 |
| 2 | 100 | 10 | 1.655 | 0.073 | 0.227 | 1.955 | 0.034 | 0.1157 | 0.0085 | 0.0579 |
| 4 | 100 | 10 | 1.871 | 0.045 | 0.327 | 2.243 | 0.001 | 0.0939 | 0 | 0.0046 |
| 6 | 100 | 10 | 2.002 | 0.038 | 0.388 | 2.428 | 0 | 0.0883 | 0 | 0.0012 |
| 10 | 100 | 10 | 2.216 | 0.034 | 0.488 | 2.738 | 0.002 | 0.1004 | 0.0001 | 0.0056 |
| Ref.[1] | 100 | 10 | - | - | - | 2.7 | - | - | - | |

We reduce the addition of artificial dissipation, considering $Sc$ = 100 and $\infty$ . Sc = $\infty$ means that artificial dissipation is not added at all. The results for $Sc$ =100 and $\infty$ are list in table 7 and table 8, respectively. For $Sc$ = $\infty$,our calculation reaches only $Wi$ = 1

Table 7. Results for increasing $Wi$ of $Re$ =100, $L$ = 100, $Sc$ =100 for Td.

| $Wi$ | $L$ | $Sc$ | $\overline{C_{dp}}$ | $\overline{C_{dv}}$ | $\overline{C_{de}}$ | $\overline{C_d}$ | $C_{lrms}$ | $St$ | $u_{rms}$ | $v_{rms}$ |
|---|---|---|---|---|---|---|---|---|---|---|
| 0 | 100 | 100 | 1.014 | 0.347 | 0 | 1.361 | 0.235 | 0.1669 | 0.0458 | 0.3091 |
| 0.5 | 100 | 100 | 1.044 | 0.303 | 0.033 | 1.380 | 0.225 | 0.1606 | 0.0477 | 0.3042 |
| 1 | 100 | 100 | 1.116 | 0.260 | 0.042 | 1.418 | 0.187 | 0.1555 | 0.0388 | 0.2584 |
| 2 | 100 | 100 | 1.229 | 0.176 | 0.063 | 1.468 | 0.045 | 0.1408 | 0.0054 | 0.0884 |
| 4 | 100 | 100 | 1.351 | 0.109 | 0.101 | 1.561 | 0.003 | 0.1305 | 0.0002 | 0.0117 |
| 6 | 100 | 100 | 1.491 | 0.092 | 0.134 | 1.717 | 0.009 | 0.1575 | 0.0009 | 0.0217 |
| 10 | 100 | 100 | 1.571 | 0.075 | 0.154 | 1.8 | 0.018 | 0.1674 | 0.0012 | 0.0334 |

As $Sc$ increases to 100, $\overline{C_d}$ is reduced to 1.8 for $Wi$ = 10, compared to $Sc$ = 10. This means that the diffusion behavior of polymer overestimates the cylinder's drag. Compared with $Sc$ = 10, $\overline{C_{de}}$ and $\overline{C_{dp}}$ reduces greatly. However, $\overline{C_{dv}}$ increases little. The addition of a small amount of artificial dissipation may weaken the concentration of elastic stress concentration area a little. However, it enlarges the concentrated area of elastic stress greatly. The instantaneous vortex distributions for for $Sc$ = 10 and $Sc$ = 100 are shown in fig. 3(a) and fig. 3(b), respectively. The distance between vortex streets on both sides of the cylinder in the vertical direction is larger, when the diffusion of polymer molecules is enhanced. The diffusion of polymer makes the scale structure of flow larger, which reduce the flow fluctuation frequency. As list in table 7, $St$ for $Wi$ = 4 and $Sc$ =100 is 0.1305, compared $St$ = 0.0939 for $Wi$ = 4 and $Sc$ =10.

Table 8. Results for increasing *Wi* of *Re* = 100, *L* = 100, *Sc* = ∞ for Td.

| $Wi$ | $L$ | $Sc$ | $\overline{C_{dp}}$ | $\overline{C_{dv}}$ | $\overline{C_{de}}$ | $\overline{C_d}$ | $C_{lrms}$ | $St$ | $u_{rms}$ | $v_{rms}$ |
|---|---|---|---|---|---|---|---|---|---|---|
| 0 | 100 | ∞ | 1.014 | 0.347 | 0 | 1.361 | 0.235 | 0.1669 | 0.0458 | 0.3091 |
| 0.5 | 100 | ∞ | 1.038 | 0.307 | 0.030 | 1.375 | 0.223 | 0.1608 | 0.0473 | 0.3031 |
| 1 | 100 | ∞ | 1.102 | 0.277 | 0.026 | 1.405 | 0.191 | 0.1560 | 0.0396 | 0.2596 |

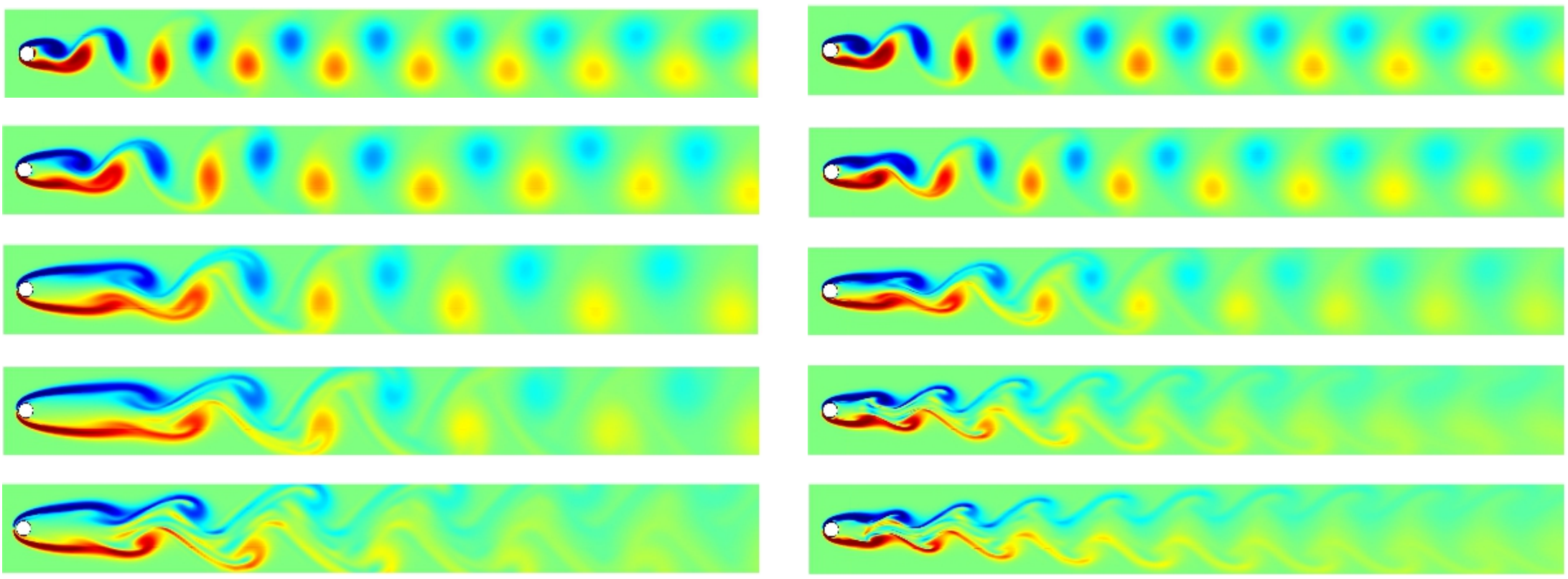

(a) *Sc* = 10, Td (b) *Sc* = 100, Td

Fig. 3. The instantaneous vortex distributions . From top to bottom are *Wi* = 1, 2, 4, 6 and 10. L is set as 100.

**B. Square reconstruction method**

Because of the instability of numerical calculation, we can only reach *Wi*=1 at Td(Sc=∞).

Table 10. Results for increasing *Wi* of *Re* = 100, *L* = 100 for Sqrt.

| $Wi$ | $L$ | $\overline{C_{dp}}$ | $\overline{C_{dv}}$ | $\overline{C_{de}}$ | $\overline{C_d}$ | $C_{lrms}$ | $St$ | $u_{rms}$ | $v_{rms}$ |
|---|---|---|---|---|---|---|---|---|---|
| 0 | 100 | 1.014 | 0.347 | 0 | 1.361 | 0.235 | 0.1669 | 0.0458 | 0.3091 |
| 1 | 100 | 1.090 | 0.276 | 0.026 | 1.392 | 0.187 | 0.1632 | 0.0343 | 0.2424 |
| 2 | 100 | 1.166 | 0.201 | 0.016 | 1.383 | 0.082 | 0.1560 | 0.0127 | 0.1360 |
| 4 | 100 | 1.220 | 0.137 | 0.011 | 1.368 | 0.015 | 0.1502 | 0.0010 | 0.0314 |
| 6 | 100 | 1.262 | 0.117 | 0.008 | 1.387 | 0.007 | 0.1595 | 0.0006 | 0.0162 |
| 10 | 100 | 1.326 | 0.107 | 0.005 | 1.438 | 0.008 | 0.1627 | 0.0009 | 0.0188 |
| 20 | 100 | 1.396 | 0.093 | 0.005 | 1.494 | 0.010 | 0.1781 | 0.007 | 0.0199 |
| 40 | 100 | 1.398 | 0.082 | 0.004 | 1.484 | 0.002 | 0.1748 | | |
| 60 | 100 | | | | 1.521 | | - | | |
| 80 | 100 | | | | 1.535 | | | | |

Table 9. Results for increasing *Wi* of *Re* = 100, *L* = 50 for Sqrt.

| $Wi$ | $L$ | $\overline{C_{dp}}$ | $\overline{C_{dv}}$ | $\overline{C_{de}}$ | $\overline{C_d}$ | $C_{lrms}$ | $St$ | $u_{rms}$ | $v_{rms}$ |
|---|---|---|---|---|---|---|---|---|---|
| 0 | 50 | 1.014 | 0.347 | 0 | 1.361 | 0.235 | 0.1669 | 0.0458 | 0.3091 |
| 1 | 50 | 1.068 | 0.283 | 0.024 | 1.375 | 0.209 | 0.1651 | 0.0365 | 0.2562 |
| 2 | 50 | 1.080 | 0.241 | 0.017 | 1.338 | 0.091 | 0.1551 | 0.0119 | 0.1438 |
| 4 | 50 | 1.066 | 0.206 | 0.012 | 1.284 | 0.022 | 0.1506 | 0.0013 | 0.0419 |
| 6 | 50 | 1.068 | 0.194 | 0.010 | 1.272 | 0.008 | 0.1433 | 0.0004 | 0.0195 |
| 10 | 50 | 1.083 | 0.187 | 0.008 | 1.278 | 0.003 | 0.1563 | 0.0002 | 0.0123 |
| 20 | 50 | 1.144 | 0.184 | 0.007 | 1.335 | 0.014 | 0.1935 | 0.0026 | 0.0304 |
| 40 | 50 | 1.158 | 0.176 | 0.004 | 1.338 | 0.016 | 0.1952 | 0.0094 | 0.0483 |
| 60 | 50 | 1.181 | 0.173 | 0.003 | 1.357 | 0.022 | 0.1922 | 0.0275 | 0.0550 |

Table 11. Results for increasing *Wi* of *Re* = 100, *L* = 200 for Sqrt.

| $Wi$ | $L$ | $\overline{C_{dp}}$ | $\overline{C_{dv}}$ | $\overline{C_{de}}$ | $\overline{C_d}$ | $C_{lrms}$ | $St$ | $u_{rms}$ | $v_{rms}$ |
|---|---|---|---|---|---|---|---|---|---|
| 0 | 200 | 1.014 | 0.347 | 0 | 1.361 | 0.235 | 0.1669 | 0.0458 | 0.3091 |
| 0.5 | 200 | 1.039 | 0.307 | 0.029 | 1.375 | 0.224 | 0.1655 | 0.0479 | 0.2983 |
| 1 | 200 | 1.094 | 0.272 | 0.025 | 1.391 | 0.188 | 0.1633 | 0.0378 | 0.2602 |
| 2 | 200 | 1.249 | 0.165 | 0.015 | 1.429 | 0.085 | 0.1551 | 0.0117 | 0.1406 |
| 4 | 200 | 1.403 | 0.075 | 0.006 | 1.484 | 0.012 | 0.1263 | 0.0006 | 0.0267 |
| 6 | 200 | 1.502 | 0.058 | 0.005 | 1.565 | 0.004 | 0.1442 | 0.0002 | 0.0115 |
| 10 | 200 | 1.667 | 0.053 | 0.004 | 1.724 | 0.007 | 0.1525 | 0.0002 | 0.0140 |
| 20 | 200 | 1.831 | 0.050 | 0.002 | 1.883 | 0.011 | 0.1613 | 0.0117 | 0.0143 |
| 40 | 200 | 1.995 | 0.060 | 0.024 | 2.079 | 0.013 | 0.1686 | 0.0171 | 0.0148 |
| 60 | 200 | 2.079 | 0.075 | 0.033 | 2.187 | 0.015 | 0.1697 | 0.0234 | 0.0172 |

## C. Logarithmic reconstruction method

Table 12. Results for increasing *Wi* of *Re* = 100, *L* = 50 for Log.

| $Wi$ | $L$ | $\overline{C_{dp}}$ | $\overline{C_{dv}}$ | $\overline{C_{de}}$ | $\overline{C_d}$ | $C_{lrms}$ | $St$ | $u_{rms}$ | $v_{rms}$ |
|---|---|---|---|---|---|---|---|---|---|
| 0 | 50 | 1.014 | 0.347 | 0 | 1.361 | 0.235 | 0.1669 | 0.0458 | 0.3091 |
| 1 | 50 | 1.078 | 0.277 | 0.027 | 1.382 | 0.189 | 0.1667 | 0.0374 | 0.2622 |
| 2 | 50 | 1.109 | 0.234 | 0.019 | 1.362 | 0.111 | 0.1575 | 0.0162 | 0.1700 |
| 4 | 50 | 1.102 | 0.196 | 0.014 | 1.312 | 0.039 | 0.1500 | 0.0028 | 0.0663 |
| 10 | 50 | 1.092 | 0.167 | 0.008 | 1.267 | 0.004 | 0.1255 | 0.0002 | 0.0084 |
| 20 | 50 | 1.116 | 0.163 | 0.006 | 1.285 | 0.012 | 0.1678 | 0.0047 | 0.0134 |
| 40 | 50 | 1.180 | 0.157 | 0.003 | 1.340 | 0.016 | 0.1814 | 0.0056 | 0.0490 |
| 80 | 50 | 1.249 | 0.128 | 0.002 | 1.379 | 0.018 | 0.1909 | 0.0147 | 0.0543 |

Table 13. Results for increasing *Wi* of *Re* = 100, *L* = 100 for Log.

| $Wi$ | $L$ | $\overline{C_{dp}}$ | $\overline{C_{dv}}$ | $\overline{C_{de}}$ | $\overline{C_d}$ | $C_{lrms}$ | $St$ | $u_{rms}$ | $v_{rms}$ |
|---|---|---|---|---|---|---|---|---|---|
| 0 | 100 | 1.014 | 0.347 | 0 | 1.361 | 0.235 | 0.1669 | 0.0458 | 0.3091 |
| 1 | 100 | 1.097 | 0.269 | 0.028 | 1.394 | 0.192 | 0.1638 | 0.0393 | 0.2700 |
| 2 | 100 | 1.205 | 0.191 | 0.018 | 1.414 | 0.112 | 0.1569 | 0.0175 | 0.1739 |
| 4 | 100 | 1.268 | 0.124 | 0.011 | 1.403 | 0.032 | 0.1439 | 0.0021 | 0.0532 |
| 10 | 100 | 1.328 | 0.087 | 0.007 | 1.422 | 0.004 | 0.1213 | 0.0001 | 0.0067 |
| 20 | 100 | 1.392 | 0.076 | 0.005 | 1.481 | 0.021 | 0.1652 | 0.0054 | 0.0144 |
| 40 | 100 | 1.458 | 0.073 | 0.003 | 1.534 | 0.023 | 0.1671 | 0.0102 | 0.0295 |
| 80 | 100 | 1.493 | 0.070 | 0.002 | 1.565 | 0.025 | 0.1690 | 0.0157 | 0.0470 |

Table 14. Results for increasing *Wi* of *Re* = 100, *L* = 200 for Log.

| $Wi$ | $L$ | $\overline{C_{dp}}$ | $\overline{C_{dv}}$ | $\overline{C_{de}}$ | $\overline{C_d}$ | $C_{lrms}$ | $St$ | $u_{rms}$ | $v_{rms}$ |
|---|---|---|---|---|---|---|---|---|---|
| 0 | 200 | 1.014 | 0.347 | 0 | 1.361 | 0.235 | 0.1669 | 0.0458 | 0.3091 |
| 1 | 200 | 1.105 | 0.265 | 0.029 | 1.399 | 0.192 | 0.1493 | 0.0387 | 0.2663 |
| 2 | 200 | 1.293 | 0.153 | 0.016 | 1.462 | 0.120 | 0.1313 | 0.0198 | 0.1777 |
| 4 | 200 | 1.467 | 0.066 | 0.007 | 1.540 | 0.0325 | 0.1200 | 0.0021 | 0.0523 |
| 6 | 200 | 1.549 | 0.053 | 0.006 | 1.608 | 0.0099 | 0.1279 | 0.0005 | 0.0176 |
| 10 | 200 | 1.656 | 0.047 | 0.004 | 1.707 | 0.0034 | 0.1096 | 0.0001 | 0.0054 |
| 20 | 200 | 1.809 | 0.046 | 0.004 | 1.859 | 0.0862 | 0.1122 | 0.0130 | 0.0163 |
| 40 | 200 | 2.103 | 0.047 | 0.003 | 2.153 | 0.1280 | 0.1152 | 0.0895 | 0.3150 |
| 80 | 200 | 2.164 | 0.048 | 0.001 | 2.213 | 0.2093 | 0.1189 | 0.1096 | 0.3925 |

### D. Summary of three stabilization techniques for long chain length

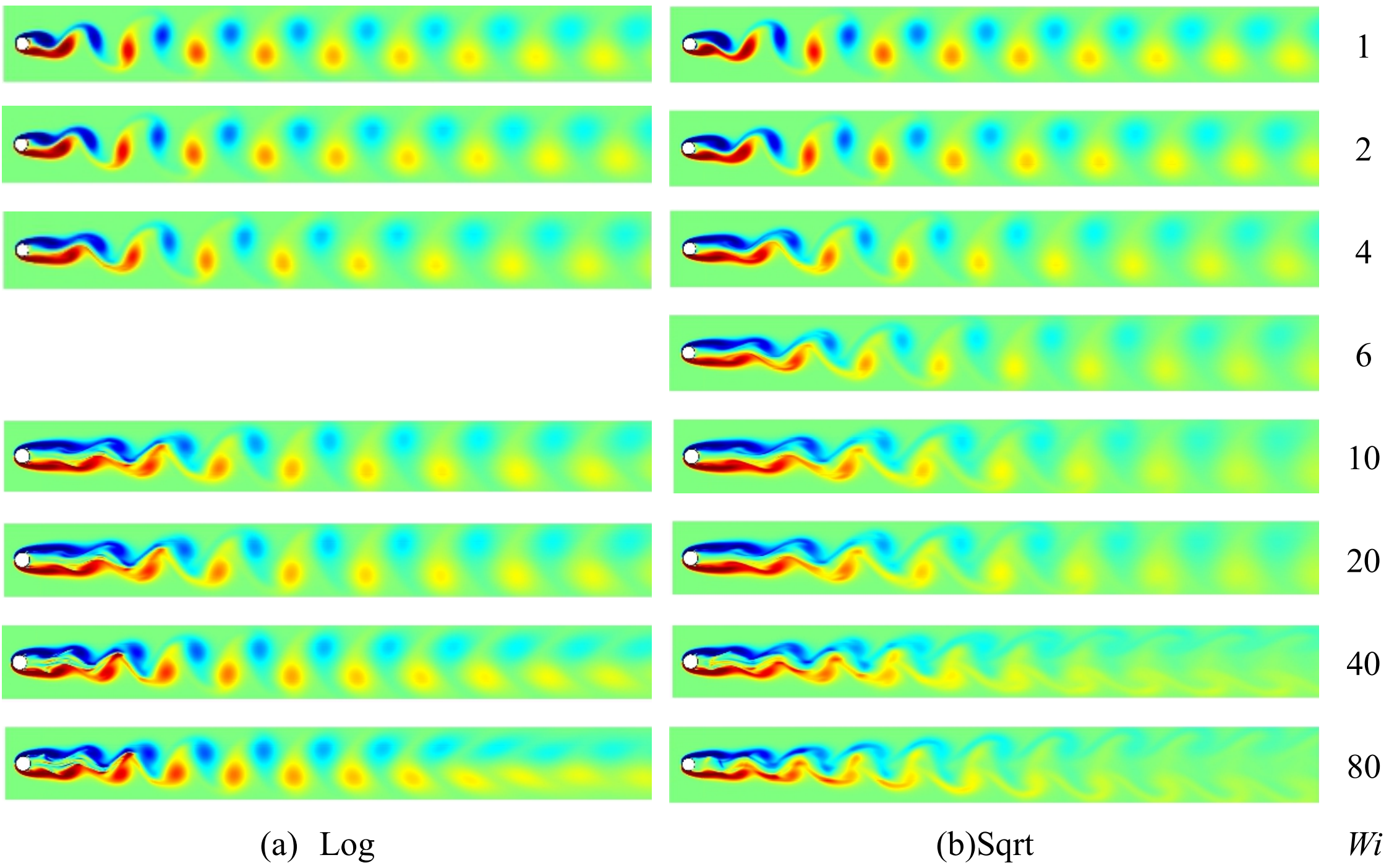

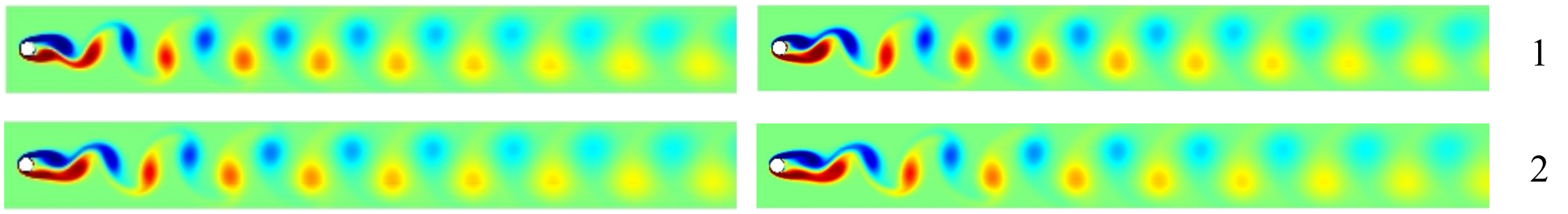

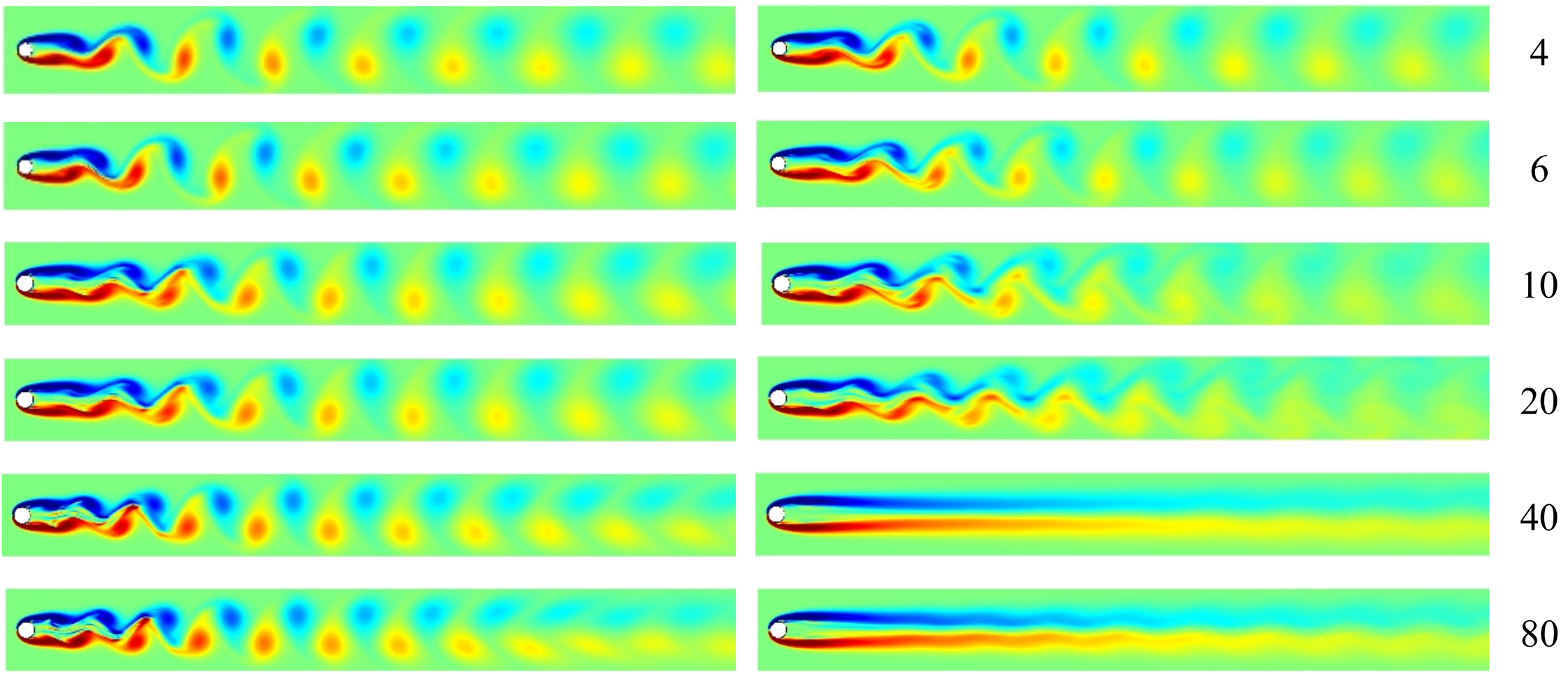

(a) Log (b)Sqrt *Wi*

Table 15.

| *Wi* | *L* | $\overline{C_d}$ | | | | | $C_{lrms}$ | | | | |
|---|---|---|---|---|---|---|---|---|---|---|---|
| Stabilization techniques | | *Sc*=10 | *Sc*=100 | *Sc*=∞ | Sqrt | Log | *Sc*=10 | *Sc*=100 | *Sc*=∞ | Sqrt | Log |
| 0 | | 1.361 | 1.361 | 1.361 | 1.361 | 1.361 | 0.235 | 0.235 | 0.235 | 0.235 | 0.235 |
| 1 | 100 | 1.699 | 1.418 | 1.405 | 1.392 | 1.394 | 0.195 | 0.187 | 0.191 | 0.188 | 0.192 |
| 2 | 100 | 1.955 | 1.468 | - | 1.383 | 1.414 | 0.034 | 0.045 | - | 0.085 | 0.112 |
| 4 | 100 | 2.243 | 1.561 | - | 1.368 | 1.403 | 0.001 | 0.003 | - | 0.012 | 0.032 |
| 6 | 100 | 2.428 | 1.717 | - | 1.387 | - | 0 | 0.009 | - | 0.004 | - |
| 10 | 100 | 2.738 | 1.8 | - | 1.438 | 1.422 | 0.002 | 0.018 | - | 0.007 | 0.004 |
| 20 | 100 | - | - | - | 1.494 | 1.481 | - | | - | 0.011 | 0.021 |
| 40 | 100 | - | - | - | 1.484 | 1.534 | - | | - | 0.013 | 0.023 |
| 80 | 100 | - | - | - | 1.535 | 1.565 | - | | - | - | 0.025 |

Table 16.

| *Wi* | *L* | $\overline{C_d}$ | | | $C_{lrms}$ | | |
|---|---|---|---|---|---|---|---|
| Stabilization techniques | | Tm | Sqrt | Log | Tm | Sqrt | Log |
| 0 | | 1.361 | 1.361 | 1.361 | 0.235 | 0.235 | 0.235 |
| 1 | 200 | - | 1.391 | 1.399 | 0.187 | 0.188 | 0.192 |
| 2 | 200 | - | 1.429 | 1.462 | - | 0.085 | 0.120 |
| 4 | 200 | - | 1.484 | 1.540 | - | 0.012 | 0.0325 |
| 6 | 200 | - | 1.565 | 1.608 | - | 0.004 | 0.0099 |
| 10 | 200 | - | 1.724 | 1.707 | - | 0.007 | 0.0034 |
| 20 | 200 | - | 1.883 | 1.859 | - | 0.011 | 0.0862 |
| 40 | 200 | - | 2.079 | 2.153 | - | 0.013 | 0.1280 |
| 80 | 200 | - | - | 2.213 | - | - | 0.2093 |

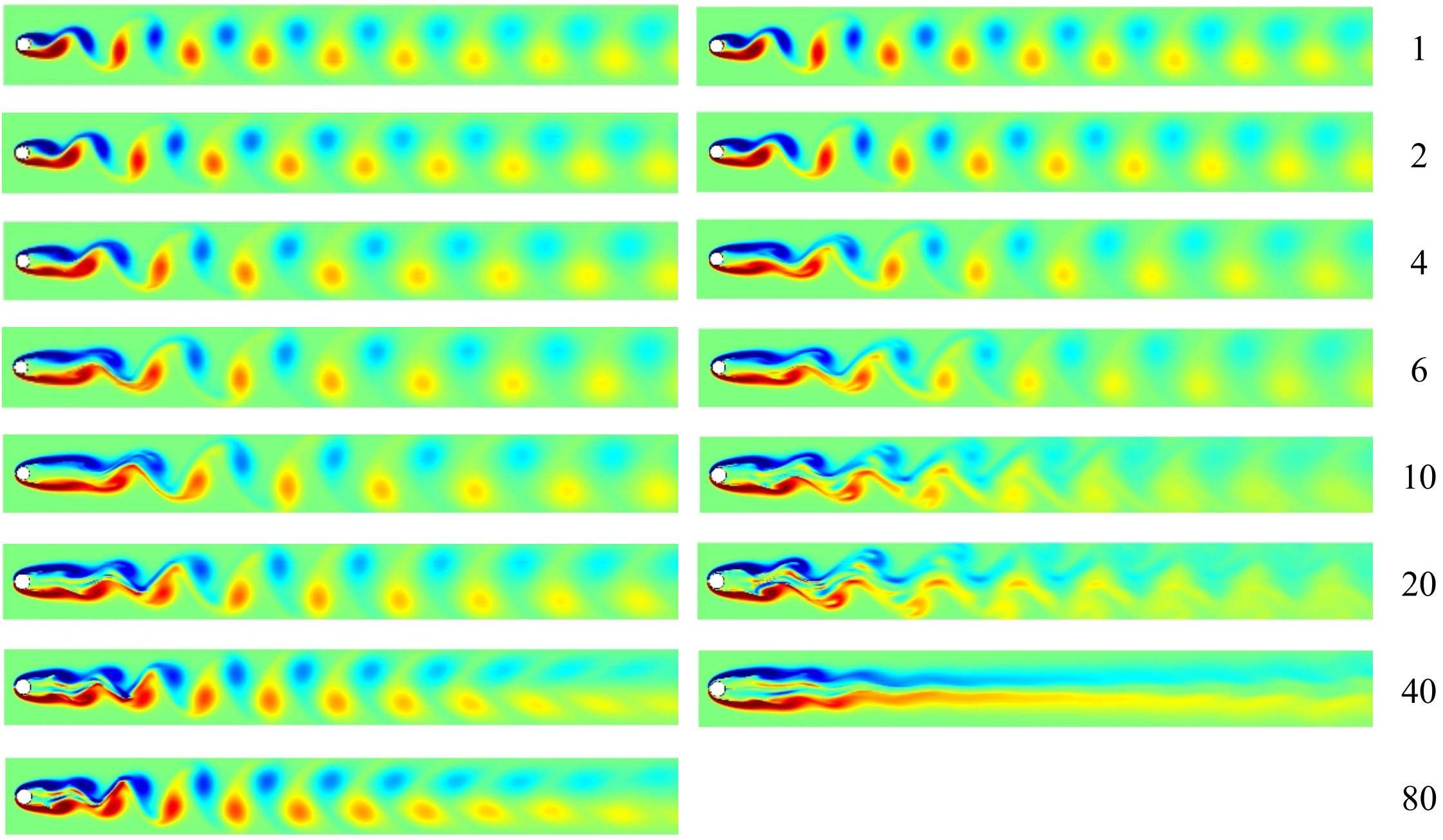

(a) Log (b)Sqrt *Wi*

## 4. Conclusion

In this study, the OpenFOAM platform, based on the finite volume method, is applied to investigate the two-dimensional viscoelastic flow past a circular cylinder. The FENE-P model, which considers the bounded elongation of polymer molecules, is chosen to describe the elastic constitutive relationship of the polymer solution. The maximum molecular chain lengths of $L$ = 10, 50, 100, and 200 are considered, which describe the molecular conformation characteristics of the polymer solution. To improve the numerical instability of the viscoelastic flow simulation, three different methods, i.e., the tranditional method (Tm) with addition of artificial viscosity, the logarithmic reconstruction method (Log), and the square root tensor method (Sqrt), are evaluated. The results show that the artificial viscosity has a little effect on the accuracy for the simulation with a small molecular chain length ($L$ = 10). However, for long molecular chain lengths such as $L$ = 100 and $L$ = 200, the addition of artificial dissipation tends to overestimate the drag, which indicates that special caution is needed to incorporate the artificial dissipation in the simulation. Moreover, the logarithmic reconstruction method shows strong grid-dependent characteristics, which may produce unphysical results.

1) Artificial dissipation has a great influence on viscoelastic flow calculation, especially on drag acted on circular cylinder.

2) Sqrt and Log are very close to the calculation of cylinder's drag. However, at high *Wi*, the influence on flow stability is quite different.